# A short-range structural insight into lithium substituted barium vanadate glasses using Raman and EPR spectroscopy as probes

## Parul Goel[1,2], Gajanan V Honnavar[2*]

[1]Dept. Of Physics, St. Joseph's College, 36 Lalbagh Road, Bangalore - 560027, India

[2]Dept. Of Physics, PES University South Campus, Bangalore - 560100, India

[*]Corresponding Author

Email: gajanan.honnavar@gmail.com

Mobile: +91-9845166113

Authors' contribution: GVH formulated and supervised the research. PG performed the experiments and collected data. PG analyzed and documented the results with inputs from GVH. PG wrote the manuscript. Both authors contributed to the manuscript.

**Abstract:**

We present a corroborative study of structural characterization of lithium substituted barium vanadate glasses using Raman and Electron Paramagnetic Resonance (EPR) spectroscopy. Investigation of the thermal and physical properties of these glasses showed a gradual increase in the concentration of non-bridging oxygen. Raman and EPR analysis gave an insight into the changing structure of the glasses. Both the spectroscopic techniques confirmed that vanadium is present in the glasses as distorted $VO_6$ octahedra. From the analysis of both spectroscopic techniques, it is proposed that the lithium ion prefers to occupy planar positions of the $VO_6$ octahedra thus reducing the tetragonal distortion and making the environment around the network forming unit in the glass matrix more homogenous as we increase lithium content. The concentration of $V^{4+}$ showed a non-monotonic variation with an increase in $Li_2O$ as indicated by Raman studies and confirmed by EPR which indicates a structural change in the distorted $VO_6$ octahedra.

## 1. **Introduction**:

There has been renewed interest in vanadium oxide and its intercalates over the last few decades due to their potential use in many technological applications especially advanced battery materials [1–4]. The unique properties of vanadium compounds are attributed to the variable oxidation states of vanadium (+3,+4,+5). In battery applications, this allows the insertion and removal of ions such as lithium in vanadium oxide compounds [5]. However, the structural framework of vanadium in any compound is decided by its coordination polyhedral with oxygen which in turn decides its stability, particularly during redox reactions[5,6]. So, it is important to understand the structure of vanadium-oxygen coordination polyhedral in any given compound and how it changes with the change in the composition of the compound. The addition of a varying amount of network modifiers in a glass compound changes its structure and properties. These short to medium range structural changes can be probed using different spectroscopic techniques as they are powerful tools for the structural characterization of glasses[7]. Raman and EPR spectroscopies are widely used to study the short range structural and coordination changes in glasses.

The existence of the octahedrally coordinated structure of vanadium in several compounds has been studied using Raman and Electron Paramagnetic Resonance (EPR) spectroscopy[8–11].

Raman is widely used to characterize the structural changes in vanadate glasses because the measured Raman peak positions and intensities are remarkably affected by the sample composition and the EPR of these glasses shows rich hyperfine structure due to the presence of vanadyl $VO^{2+}$ ions. These techniques were able to estimate the changes in the specific coordination of network former($V_2O_5$) and estimate bond lengths up to a few angstroms[12,13].

Both these non-destructive spectroscopic techniques have been widely used to determine the structural changes in binary and ternary vanadate glasses[14–18]. Specific studies have explored the effect of variation in $Li^+$ ion concentration on the structure of vanadate glasses using these techniques[14,19,20]. However, vanadate glasses with a combination of alkaline-earth oxides and alkali oxides have not been extensively studied. Also, in most of the studies, these spectroscopic techniques are used as stand-alone characterization methods. In the few reports where a combination of these techniques is used, the results of one technique are not used to support and enhance the observations from the other[17,21,22].

Our present work aims at studying the structural changes in $V_2O_5$- $BaO$-$Li_2O$ glasses as a function of increasing the concentration of $Li_2O$. In our samples, we have kept the TMO content constant and varied the amount of $Li_2O$ at the expense of BaO. A detailed study of structural changes in these types of oxide glasses with constant TMO content and a variable amount of alkaline-earth oxide and alkali metal oxide has not been done so far. We were able to probe the short-range structural changes in these glasses via a corroborative study of Raman and EPR spectroscopy.

## 2. **Experimental Studies** :

Glasses with $60V_2O_5$- $(40-x)$ $BaO$-$xLi_2O$ where x=10,15,20,25(mol%) were prepared using high purity $V_2O_5$ and $Li_2CO_3$ procured from Sigma Aldrich and $BaCO_3$ from Alfa Aesar. Homogenous samples weighing 5g were synthesized using the conventional melt-quench technique [23]. The starting materials weighed in appropriate proportion were put in porcelain crucibles and melted in a muffle furnace at 800°C for 30 min. The melt was then quenched between brass plates to prepare glass. The samples were then annealed at 100°C for 8 hours and left to cool down slowly to room temperature to relieve any mechanical stress.

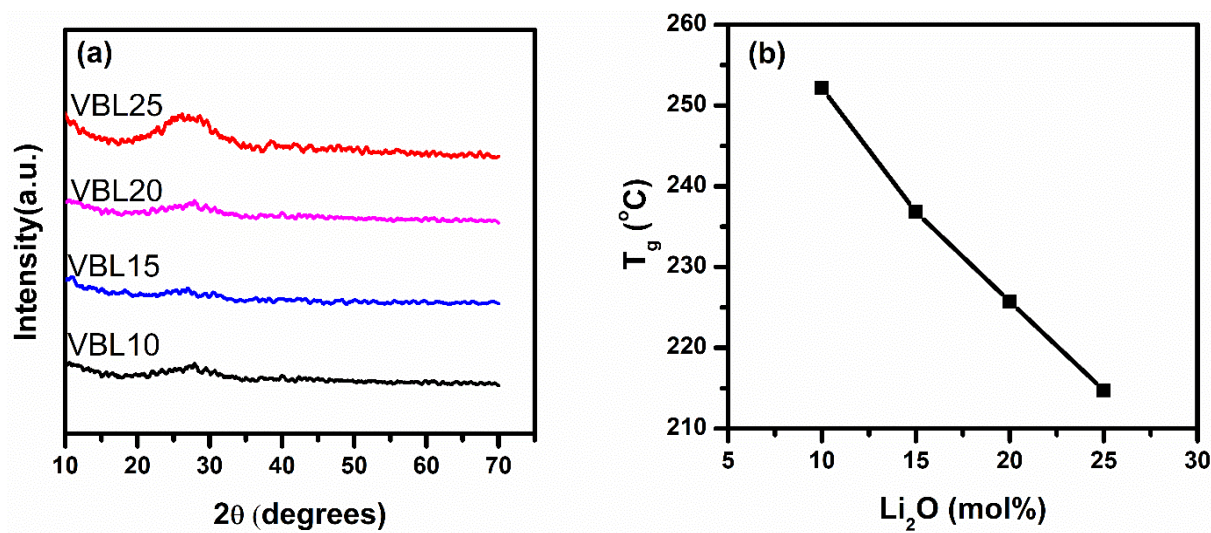

*Figure 1(a) XRD pattern of the prepared glasses showing the absence of any sharp crystallization peaks which confirms the amorphous nature. (b) Variation in glass transition temperature ($T_g$) for all glass samples.*

The glass transition temperature ($T_g$) was measured using Differential Scanning Calorimeter (DSC 2920 TA Instruments) by heating the samples at a constant rate of 5°C/min. with an accuracy of $\pm\ 0.05°C$. (The cumulative DSC graph is given in **supporting material**).

The density of prepared samples 'ρ' was measured using the Archimedes principle with Xylene as the immersion liquid. The density of Xylene used was 0.861 g/cm³ as mentioned by the manufacturer. All measurements were done for 3 different sets of samples to take care of the random errors. The molar volume and the error in molar volume for each glass sample were calculated from the density data.

The average cross-link density gives a quantitative analysis of thermal properties in terms of structural parameters. So, these properties were further investigated using the average cross-link density of the glasses ($\bar{n}_c$). Average cross-link density can be calculated from few physical parameters but the variation in it gives an insight into the variation in glass transition temperature[24]. The calculated values of molar volume and $\bar{n}_c$ are tabulated in **Table 1**. (Details of calculation of molar volume and average cross-link density are given in **supporting material**)

The Raman spectra were recorded at room temperature using a LabRAM HR Raman spectrometer from Holmarc with a CCD camera. The spectral resolution of the instrument was 0.4 cm$^{-1}$. The data was obtained in the spectral range from 200-1100 cm$^{-1}$. This data was then baseline corrected and smoothened to 100 points using the Shavitzky-Golay method. The temperature and frequency dependence of the Raman scattered data was then removed to obtain a reduced Raman intensity $R(\omega)$ by applying temperature-frequency correction[25]

$$R(\omega) = I(\omega)(\omega_0 - \omega_i)^{-4}\omega_i B \qquad (1)$$

Here $I(\omega)$ is the measured experimental Raman intensity, $\omega_0$ is the frequency of laser excitation line in cm$^{-1}$, $\omega_i$ is the Raman shift in cm$^{-1}$ and $B$ is the temperature factor given by Bose-Einstein formula as $B = 1 - \exp\left(-\frac{hc\omega_i}{k_B T}\right)$ (where $T$ is the room temperature in K). The spectral intensity was then scaled between 0 and 1 to obtain the final reduced intensity. The data thus obtained was again corrected for baseline and de-convolved using Gaussian line shape between the spectral range of 700-1100 cm$^{-1}$ to obtain the various V-O bond stretching vibrations for structural analysis of the various glass samples as the short-range structure of the glasses is reflected by a series of bands in this high frequency region.

The EPR spectra of powdered samples were obtained using JEOL, JES-FA200 ESR spectrometer operating in 9.4 GHz range in the X-band (measured with an accuracy of 10$^{-6}$

GHz) with a modulation frequency of 100 kHz. The spectra for these glasses were obtained at room temperature. The magnetic field was swept with a width of ± 150mT with 350mT as the center field. The experimental error involved in measurement was ± 0.0001mT. Each of the samples was weighed and the intensity of each sample was normalized by the weight of the corresponding sample.

## 3. Results and Discussion:

### 3.1 Thermal properties:

From the DSC of all the studied glass samples, we determined the glass transition temperature ($T_g$) and the values obtained are listed in **Table1.** $T_g$ shows a monotonic decrease from 252.2 °C in the glass with 10 mol% $Li_2O$ to 214.7 °C in the glass with 25 mol% $Li_2O$ as barium oxide is substituted by lithium oxide (as shown in **Figure 1(b)**). The gradual increase in $Li_2O$ content, which is more ionic compared to both $V_2O_5$ and BaO tends to create more non-bridging oxygens (NBOs) which softens the glass network.

### 3.2 Density and Molar Volume:

A perusal of **Table 1** reveals that both density and molar volume decrease linearly as the $Li_2O$ content increases in the present glass system. A decrease in density may be attributed to the lowering of the total molecular weight of the glass because a lighter component ($Li_2O$, molecular weight =29.87 g/mol) replaces a heavier component (Barium oxide 153.32 g/mol). The increase in the NBOs causes a decrease in the molar volume and the smaller ionic radius of $Li^+$ (76 pm) as compared to $Ba^+$ (138 pm) also contributes to this decrease.

The average cross-link density showed a decrease from 3.76 to 3.46 with $Li_2O$ content increasing from 10 mol% to 25 mol% as shown in **Table 1**. The decrease in the average cross-link density correlates with the decrease in T$_g$ [24] and it also indicates the creation of more NBOs.

Both the thermal and physical analysis of the glass system under study has hinted at an increase in NBOs. This could result in structural changes in the glasses. To investigate the short to medium range structural changes in these glasses, we carried out Raman and EPR analysis.

**Table1**: Composition, density, molar volume, glass transition temperature (T$_g$), the concentration of Li and V ions and the average cross-link density of prepared glass compositions.

| Sample name | Composition (mol%) V$_2$O$_5$  BaO  Li$_2$O | T$_g$ (±0.05) (°C) | Density ($\rho \pm d\rho$) (g/cm$^3$) | Molar Volume ($V_m \pm dV_m$) (cm$^3$) | n(V) (x10$^{22}$ cm$^3$) | $\bar{n}_c$ |
|---|---|---|---|---|---|---|
| VBL10 | 60   30   10 | 252.2 | 3.494±0.014 | 45.243±0.183 | 1.600 | 3.76 |
| VBL15 | 60   25   15 | 236.8 | 3.413±0.004 | 44.508±0.052 | 1.624 | 3.66 |
| VBL20 | 60   20   20 | 225.7 | 3.302±0.012 | 44.135±0.160 | 1.638 | 3.56 |
| VBL25 | 60   15   25 | 214.7 | 3.197±0.006 | 43.650±0.082 | 1.656 | 3.46 |

\* T$_g$ for each sample listed here was measured at a scan rate of 5°C/min.

### 3.3 Raman Spectroscopy

We have fitted Gaussian peaks to the spectra in the 700-1100 cm$^{-1}$ range to a set of 5 bands, out of which four bands correspond to V-O bond stretching vibrations which are denoted by *Pi* where (*i*=1,2,3,4). Peaks below 800 cm$^{-1}$ correspond to bending vibrations and will not be discussed here. The Raman spectra and the deconvolution of various V-O bonds along with a sketch of the VO$_6$ polyhedral structure that the V$_2$O$_5$ assumes in these types of glasses are shown in **Figure 2**. The various V-O bond vibrations suggest that vanadium in these glasses is present in the form of distorted VO$_6$ octahedra[8,26]. The fitted peaks are attributed according to Hardcastle et al.[8]: the highest frequency band at 991 cm$^{-1}$ (*P*1 band) is due to V$^{5+}$=O bond stretching vibrations[8,9], the next band at 961 cm$^{-1}$ (*P*2 band) is assigned to stretching vibrations of $V^{4+}=O$ bond[9,19,26], the next band around 925 cm$^{-1}$ (*P*3 band) is for O-V-O stretching vibrations within the $VO_6$ structure[26] and the vibration band at 830 cm$^{-1}$ (*P*4 band) is due to bridging asymmetric stretching vibrations of V-O-V[26] that extend over to the nearest neighboring $VO_6$ polyhedron (as shown in **Figure 2c**).

The parameters like peak position, area under the peak and Full Width at Half Maximum (FWHM) along with their uncertainties were extracted from the well-fitted spectra. (Rest of the fitted graphs are provided in the **supporting material**.)

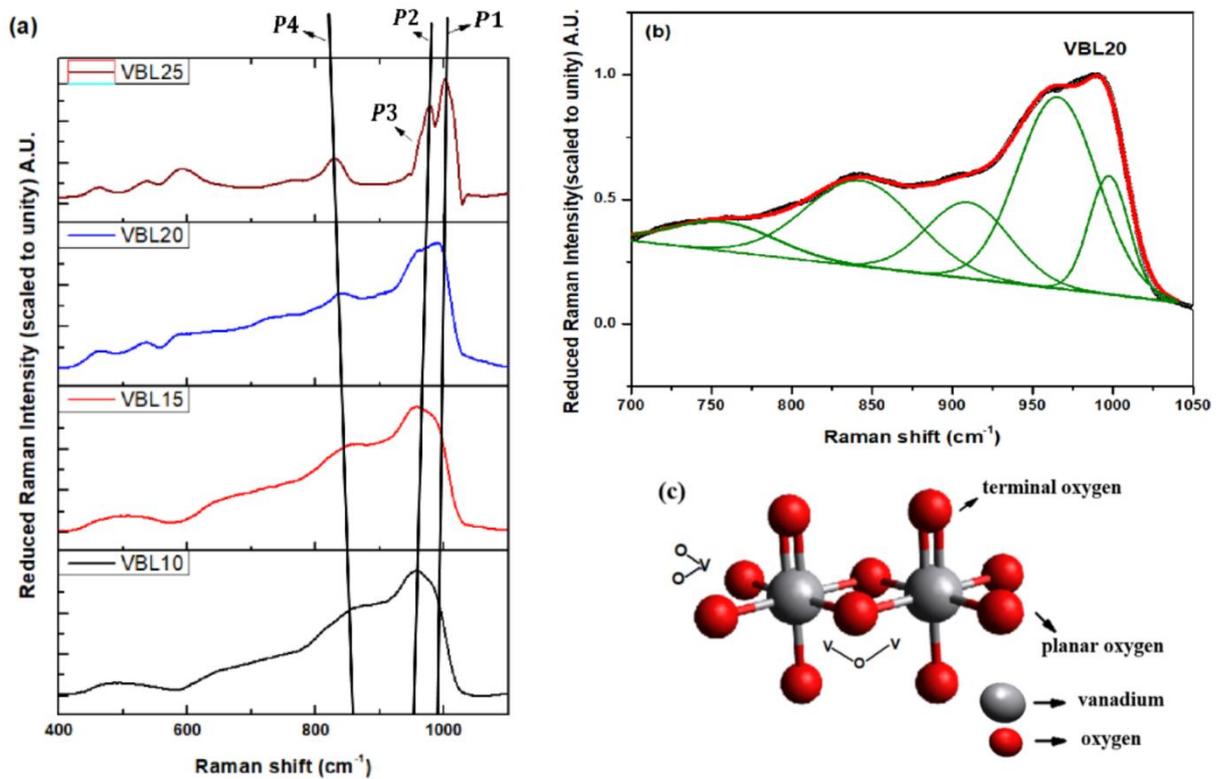

*Figure 2 (a) Reduced Raman spectra of all the glass samples. (b) Fitted Raman spectra with deconvoluted Raman peaks for a representative glass sample. (c) Sketch of VO$_6$ polyhedron in V$_2$O$_5$ based glasses.*

The graph representing peak positions along with their uncertainties is plotted in **Figure 3.** The uncertainties in Raman peak shifts vary from 0.4cm$^{-1}$ to 1.2cm$^{-1}$. The peak *P*4 shows a large red-shift as we increase the Li$_2$O content. *P*4 corresponds to V-O-V bridging vibration between the two adjacent VO$_6$ octahedra. A shift in the Raman peak also gives a measure of bond length variation according to equation (2) below. With the addition of $Li_2O$ into the glasses, some $Li^+$ ions can break the bridging oxygens in the V-O-V bond, forming the non-bridging oxygens. As $Li^+$ ion dangles to this $O^-$ ion forming $V-O^--Li^+$ bond, the $Li^+$ ion pulls the $O^-$ ion due to its highly electropositive nature, stretching the attached V-O bond[15] which decreases the vibrational frequency of the V-O-V bond.

Bands $P$1 and $P$2 are blue-shifted with a smaller magnitude (~13-14 $cm^{-1}$) as compared to the shift in $P$4 (~30 $cm^{-1}$). Since V=O is a strong, polar bond, the variation in the surrounding environment affects it only a little. As the V-O-V bond elongates, it causes some compression of the polar V=O bond length. Also, as some of the $Li^+$ replace the bigger $Ba^{2+}$ in the interstitial sites, the structure collapses a little, which also causes a decrease in the V=O bond length and an increase in the Raman frequency. Hence, both $P$1 and $P$2 show a shift to high frequencies.

$P$3 corresponds to the Raman peak of O-V-O bonds which do not take part in the glass network formation and show a non-monotonic variation. The decrease in the frequency up to VBL20 corresponds to an increase in the length of this bond. However, this bond length increases from VBL20 to VBL25, which could be due to a change in the surrounding environment of $VO_6$ octahedra as $Li^+$ is being substituted.

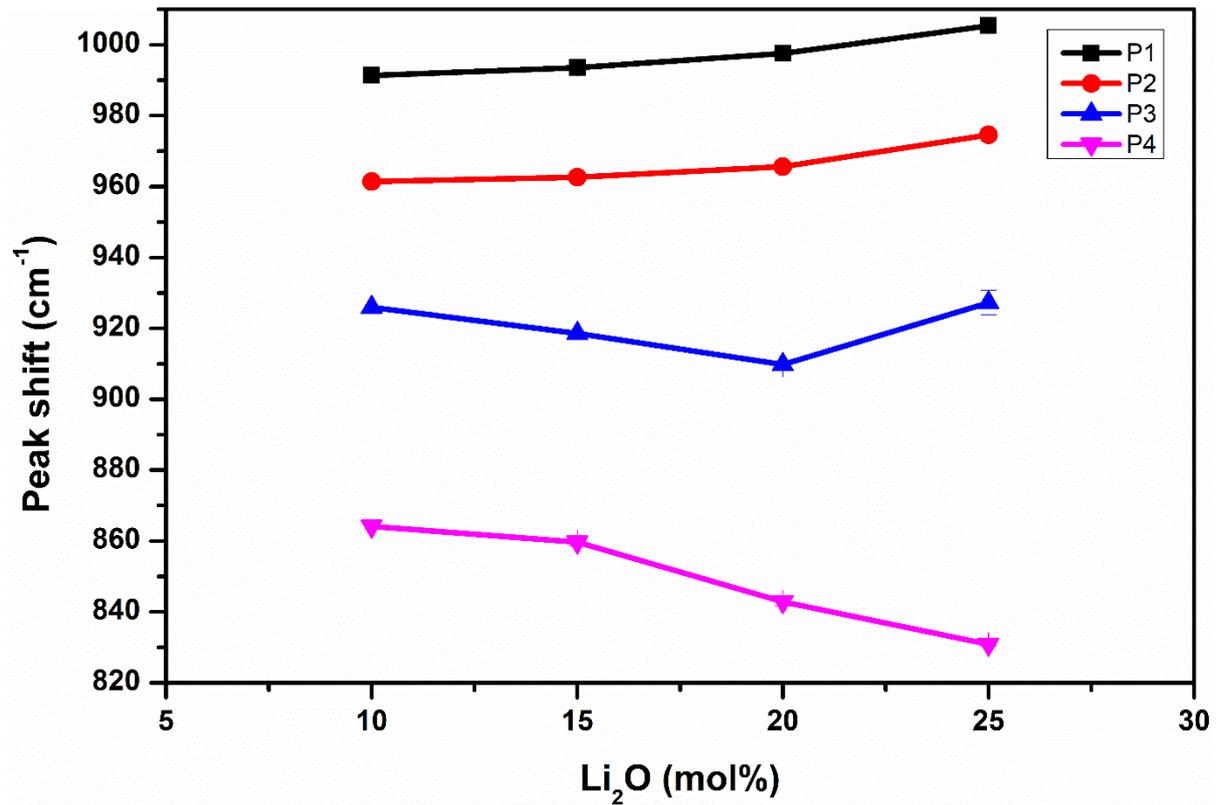

*Figure 3 Variation in peak shift with varying $Li_2O$.*

The area under each peak shows the number of species corresponding to that particular bond. **Figure 4(a)** shows the area under each of the peaks for different glass compositions. The uncertainties in peak area vary from 0.4cm$^{-1}$- 2.3cm$^{-1}$. The peak area of *P*2 increases non-monotonically. It increases from VBL10 up to VBL20 but decreases from VBL20 to VBL25 which indicates that the number of V$^{4+}$ ions forming V$^{4+}$=O bond should also show similar non-monotonic variation in the glasses. This is later confirmed by observations from EPR spectra (as shown in **Figure 7(a)**). The peak area of *P*1 representing the number of V$^{5+}$ ions in each glass showed a complimentary variation to *P*2. This shows that most of the V$^{4+}$ are created by direct conversion from V$^{5+}$. Also, since the V$_2$O$_5$ content in our glass compositions is kept constant, so the decrease in V$^{4+}$ ions from VBL20 to VBL25 indicates a structural change in the distorted VO$_6$ octahedra.

The peak area of *P*4 shows a continuous decrease which confirms that more and more NBOs are formed as we increase the concentration of Li$_2$O. The peak area of *P*3 shows small variations as the number of O-V-O species does not change much.

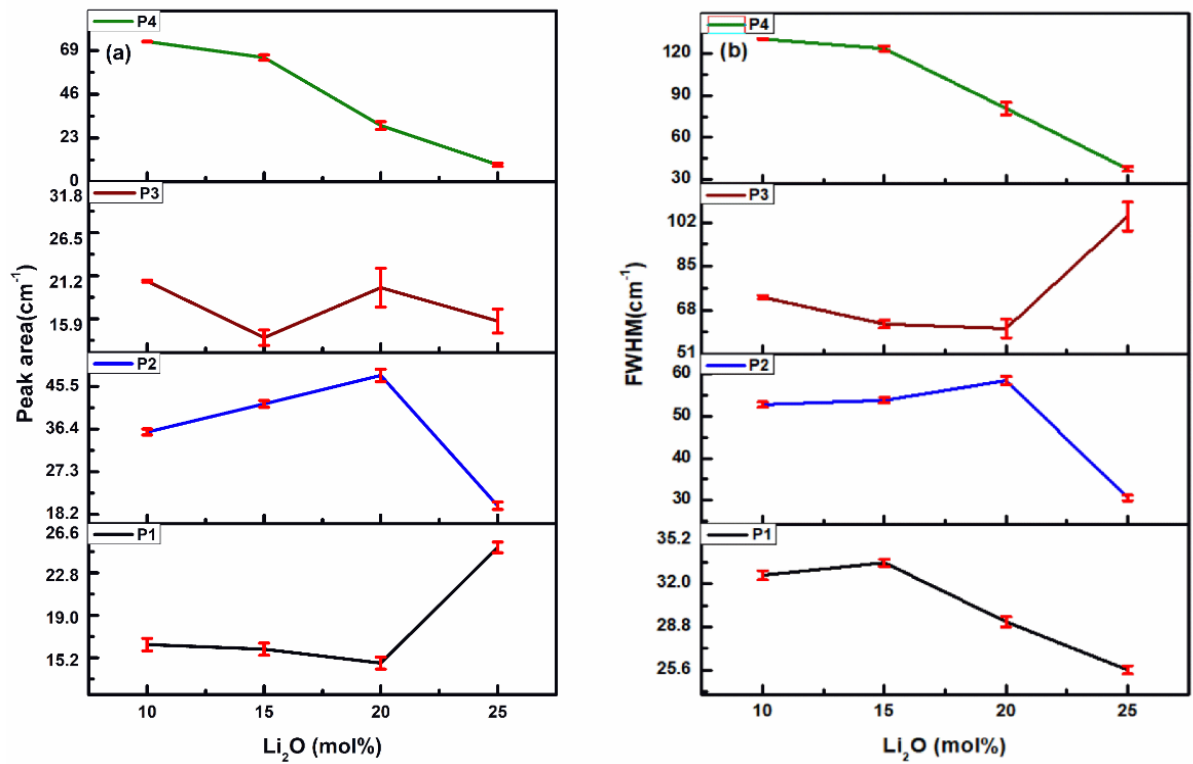

*Figure 4(a) Variation in peak area with varying Li$_2$O content. (b) Variation in FWHM with varying Li$_2$O content.*

**Figure 4(b)** shows the variation in Full-Width at Half Maximum (FWHM). The uncertainties in the value of FWHM vary from 0.4cm$^{-1}$- 4.2cm$^{-1}$. The width of each peak as indicated by FWHM gives information about the distribution of a particular bond length. *P*1 and *P*2 show non-monotonic variation in FWHM. *P*4 shows a much larger decrease in FWHM because the substituted Li$^+$ ion is smaller in size to that of Ba$^+$ but similar in size to that of vanadium ion. So, lithium substitution makes the environment around this bond more homogenous. Thus, the V-O-V bond becomes more and more homogenous. The FWHM of the O-V-O bond (*P*3) shows a non-monotonic variation where the bond becomes more homogenous up to VBL20 but then the FWHM increases which shows that the bond environment becomes non-homogeneous.

The V-O stretching frequencies ($\nu$) can be related to their bond lengths ($R$) using the exponential relation as applied by Hardcastle[8]:

$$\nu = A \exp(BR) \qquad (2)$$

where $A$ and $B$ are fitting parameters. The best fit to the experimental values was obtained with the values of $A$ and $B$ as 21349 and -1.9176 respectively[8]. So, the above equation becomes:

$$\nu = 21349 \exp(-1.9176R) \qquad (3)$$

Equation (3) gives $R$ in Å if $\nu$ is in cm$^{-1}$. Using equation (3), the most probable bond length can be obtained. These values for V-O-V are summarized in **Table 2**. As can be seen from the observed values, the bond length keeps increasing with increasing lithium concentration. This type of estimation of bond lengths was previously performed by Gajanan et al.[12] and was supported by the DFT simulations by Saetova et al. and Saiko et al.[27,28]

**Table 2** Approximate V-O-V bond length as calculated from de-convoluted frequencies around 850 cm$^{-1}$.

| Composition | Most probable bond length (Å) |
|---|---|
| VBL10 | 1.673 |
| VBL15 | 1.676 |
| VBL20 | 1.686 |
| VBL25 | 1.693 |

### 3.4 EPR Spectroscopy

The EPR spectra in glasses were first evaluated by Sands[29]. Siegel and many others studied the spectra of VO$^{2+}$ in glasses[11,16,30]. They considered that V$_2$O$_5$ is present in the glasses as VO$_6$ octahedra which is tetragonally distorted as shown in **Figure 5(b)**.

The obtained EPR spectrum is attributed to the presence of $V^{4+}$ in the glass samples as this species of vanadium is EPR active due to the presence of a single unpaired electron[13]. $^{51}V$ in $V^{4+}$ has an outer shell electronic configuration of $3p^6 3d^1$. So, in vanadyl ion ($VO^{2+}$) with electron spin S=1/2 and nuclear spin I=7/2, the total number of hyperfine components are (2I+1) and its rich hyperfine structure consists of 8 lines. However, due to g anisotropy, some of the components overlap and hence not all the lines are well resolved in the observed spectra. Corresponding to the various $m$ values (like ±1/2, ±3/2, ±5/2 and ±7/2), the magnetic field values were obtained from the first derivative spectra (shown in **Figure 5(a)**). The values of $g_{||}$ and $g_\perp$ were evaluated from the spin Hamiltonian. (Rest of the spectra are provided in the **supporting material**.)

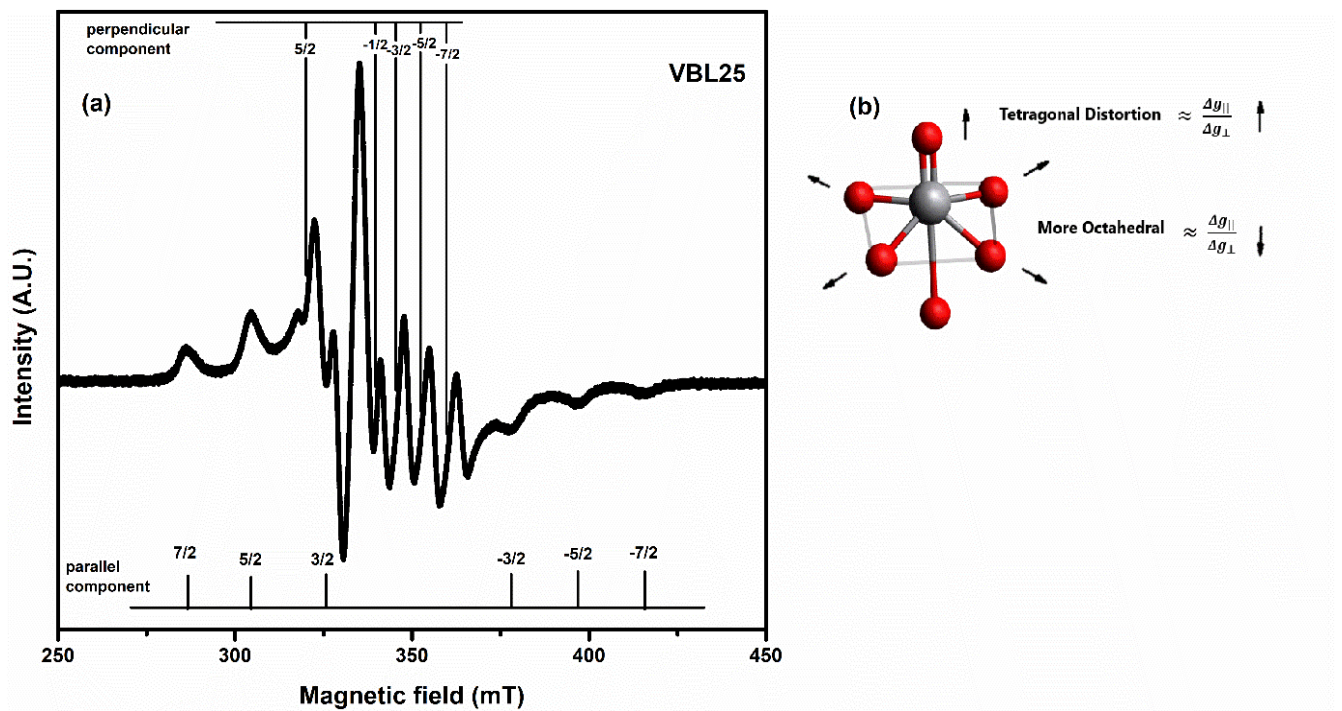

*Figure 5(a) Representative EPR spectrum of a glass sample. Positions of parallel and perpendicular components of the spectrum identified for evaluating $g_{||}$ and $g_\perp$ are marked in the figure. (perpendicular +1/2 and +3/2 components are not shown as they were not used in calculation) (b) Sketch of $VO_6$ octahedra with tetragonal distortion.*

The spin Hamiltonian of the axially symmetric field can be used to evaluate the spectra of these glasses. It is evaluated to obtain the components of $g$ tensor and hyperfine coupling tensor $A$ and is given as[16]

$$H = g_{\|}\beta H_z S_z + g_{\perp}\beta(H_x S_x + H_y S_y) + A_{\|} S_z I_z + A_{\perp}(S_x I_x + S_y I_y) \quad (4)$$

where $\beta$ is Bohr magneton, $g_{\|}$ and $g_{\perp}$ are the parallel and perpendicular principal components of the $g$ tensor. $A_{\|}$ and $A_{\perp}$ are the parallel and perpendicular principal components of the hyperfine coupling constant. $S_x, S_y, S_z$ and $I_x, I_y$ and $I_z$ are the components of the spin operator of electron and nucleus respectively and $H_x, H_y$ and $H_z$ are the components of the magnetic field. Magnetic field positions of the eight hyperfine components are given by the solution of the Hamiltonian in equation (4) as[31]

$$H_{\|}(m) = H_{\|}(0) - A_{\|}m - \frac{A_{\perp}^2}{2H_{\|}(0)}\left(\frac{63}{4} - m^2\right) \quad (5)$$

$$H_{\perp}(m) = H_{\perp}(0) - A_{\perp}m - \frac{(A_{\|}^2 + A_{\perp}^2)}{4H_{\perp}(0)}\left(\frac{63}{4} - m^2\right) \quad (6)$$

Here $m$ is the nuclear magnetic quantum number, $H_{\|}(0)$ and $H_{\perp}(0)$ are the magnetic fields corresponding to $m = 0$. $H_{\|}(0) = \frac{h\nu}{g_{\|}\beta}$ and $H_{\perp}(0) = \frac{h\nu}{g_{\perp}\beta}$, where $h$ is Plank's constant and $\nu$ is the spectrometer frequency.

From the observed first derivative (experimental) spectra, the field positions for various EPR transitions corresponding to different $m$ values were obtained. The field positions for parallel values given by equation (5) were obtained by locating broad peaks in the derivative spectrum and for perpendicular values, the field positions in equation (6) approximate best to the mid-points of the individual lines. The procedure followed in locating these components is similar to that followed by Muncaster et al., Bandyopadhyay and McKnight et al.[20,31,32].

The first approximation to the values of $H_{\parallel}(0), H_{\perp}(0), A_{\parallel}$ and $A_{\perp}$ were obtained from the least-square straight-line fits to equations:

$$H_{\parallel}(m) = H_{\parallel}(0) - A_{\parallel}m \qquad (7)$$

$$H_{\perp}(m) = H_{\perp}(0) - A_{\perp}m \qquad (8)$$

The values of $H_{\parallel}(0)$ and $H_{\perp}(0)$ were obtained from the intercepts and $A_{\parallel}$ and $A_{\perp}$ were obtained from the negative slopes of the linear fits to equations (7) and (8). The second order terms of $m$ in equations (5) and (6) were calculated by applying these initial values. Then by substituting the values of these second order terms in (5) and (6), the final values of $H_{\parallel}(0), H_{\perp}(0), A_{\parallel}$ and $A_{\perp}$ were obtained from another straight-line fit, similar to the method followed by Muncaster et al. and McKnight et al.[20,31]. (The detailed calculations are shown in **supporting material**).

The parallel components with $m = -\frac{7}{2}, -\frac{5}{2}, -\frac{3}{2}, \frac{3}{2}, \frac{5}{2}, \frac{7}{2}$ were well resolved in all spectra and used for determining the parameters in equation (5). All the perpendicular components except $m = \frac{1}{2}, \frac{3}{2}$ were used in determining the parameters of equation (6). As $H_{\perp}\left(\frac{1}{2}\right)$ and $H_{\perp}\left(\frac{3}{2}\right)$ components are least anisotropic, so the difference between the observed and calculated values is high[31]. So, they were constantly found to add greater error in the fitted parameters and hence were excluded from the calculations. Values of various obtained parameters are listed in **Table3**.

For the vanadyl $VO^{2+}$ to exist in tetragonally distorted octahedral symmetry in the glasses, the conditions, $g_{\parallel} < g_{\perp} < g_e$ and $A_{\parallel} > A_{\perp}$ should be satisfied[33]. A perusal of **Table 3** shows that these conditions are satisfied in our glass samples. Here, $g_e$ is free electron $g$ parameter with value 2.0023.

**Table 3** EPR Spin Hamiltonian parameters

| Composition | $g_{\|\|}$ | $g_\perp$ | $A_{\|\|}$ (x10$^{-4}$ cm$^{-1}$) | $A_\perp$ (x10$^{-4}$ cm$^{-1}$) | $\dfrac{\Delta g_{\|\|}}{\Delta g_\perp}$ |
|---|---|---|---|---|---|
| VBL10 | 1.9200±0.0012 | 1.9762±0.0013 | 167.99±0.74 | 59.14±1.09 | 3.1533 |
| VBL15 | 1.9188±0.0034 | 1.9756±0.0015 | 167.15±2.22 | 58.49±0.91 | 3.1273 |
| VBL20 | 1.9188±0.0030 | 1.9748±0.0013 | 171.73±1.95 | 59.70±1.20 | 3.0364 |
| VBL25 | 1.9171±0.0033 | 1.9735±0.0016 | 166.59±2.18 | 58.67±1.15 | 2.9653 |

The ratio $\dfrac{\Delta g_{\|\|}}{\Delta g_\perp}$, where $\Delta g_{\|\|} = g_e - g_{\|\|}$ and $\Delta g_\perp = g_e - g_\perp$, is indicative of the tetragonal distortion of the octahedral crystal field[31]. A larger tetragonal distortion results in a greater value of this ratio and vice-versa[13].

**Figure7(b)** shows the variation of $\dfrac{\Delta g_{\|\|}}{\Delta g_\perp}$ with varying lithium concentration. The ratio decreases monotonically as the Li content in the glass increases. This indicates that the VO$_6$ polyhedron tends towards more and more octahedral symmetry. From this observation, we can argue that the substituted lithium has a larger tendency to occupy the planar positions than the terminal positions. Due to this $\Delta g_\perp$ value increases and the ratio decreases. The prolate structure of VO$_6$ becomes more and more spherically symmetric. This in turn supports the observation from the Raman study that the structure is more homogeneous and the substitution of lithium is preferred at the planar positions over the terminal positions.

The concentration of the total number of vanadium ions (given in **Table1**) was obtained from density data (The detailed calculation is given in **supporting material**). The total number of V$^{4+}$ species was obtained by finding the area under the normalized intensity spectrum by doubly integrating it[13]. **Figure 7(a)** shows that while the total number of vanadium ions increases slightly with an increase in Li$_2$O concentration (though TMO content is constant) but the normalized number of V$^{4+}$ species (obtained by dividing the total number of V$^{4+}$ ions by the total number of vanadium ions) shows non-monotonic variation. It increases up to VBL20 and then decreases from VBL20 to VBL25. This observation supports the non-monotonic variation in the peak area of $V^{4+}$=O as obtained from Raman spectra. This shows that the environment around V=O changes from VBL20 to VBL25 due to an increase in $Li^+$ ions. As we increase the concentration of $Li_2O$, the number of NBOs increases but with a large concentration of $Li_2O$, more lithium atoms may occupy positions near to the V=O bond as shown in **Figure 6**. Since lithium is electropositive, so its presence close to the V=O bond pulls the electron towards oxygen. This increases the concentration of $V^{5+}$ in VBL25 glass, thereby reducing the concentration of $V^{4+}$ from VBL20 to VBL25.

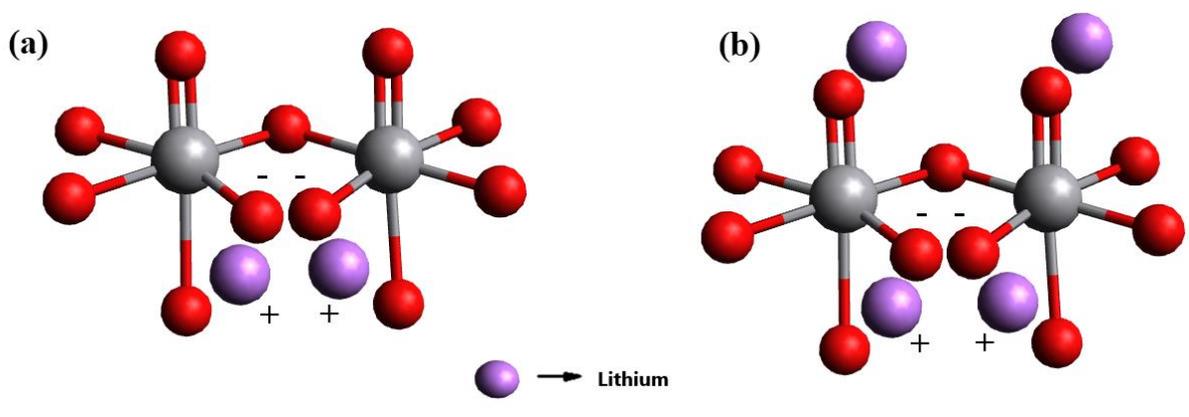

*Figure 6(a) Sketch showing the formation of non-bridging oxygens on the addition of $Li_2O$ (low concentration) in the glass. (b) at high concentration of $Li_2O$.*

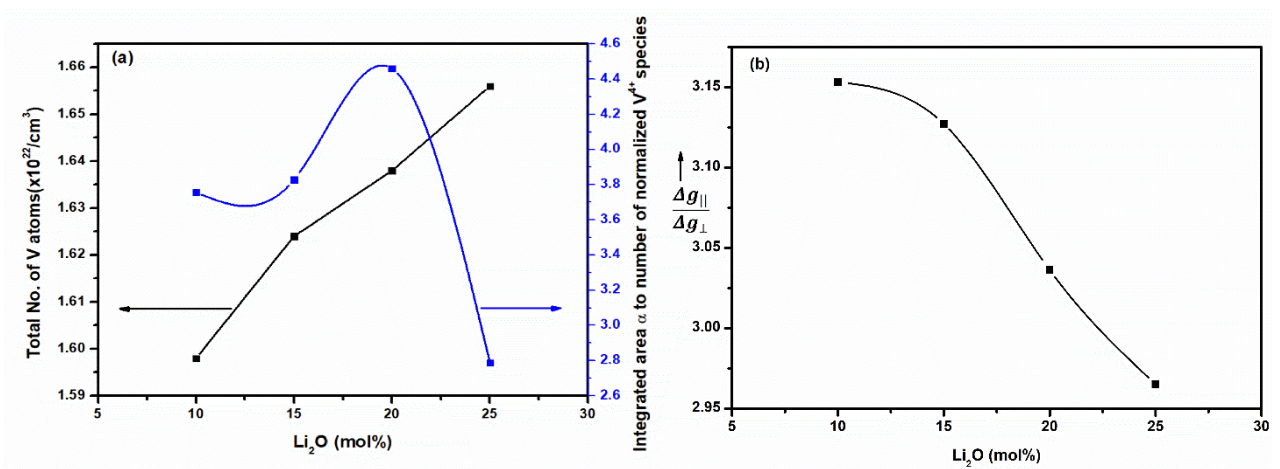

*Figure 7(a) Variation of concentration of V and the number of normalized $V^{4+}$ species with variation in Li$_2$O concentration for all glass samples. (b) Variation of $\frac{\Delta g_{\parallel}}{\Delta g_{\perp}}$ with Li$_2$O concentration. (lines are a guide to the eye).*

## 4. Conclusions:

In this communication, we have evaluated the effect of systematic substitution of the lithium ion on the structure of barium vanadate glasses.

The drop in the glass transition temperature with a rise in Li$_2$O content in the glasses at the expense of BaO relates directly to the continuous increase in the number of NBOs and the decrease in the average cross-link density. The collapsing skeletal structure as indicated by the decreasing molar volume also indicates an increase in the non-bridging oxygens.

Raman analysis which is a non-destructive tool to probe the structure supports the above observation. The gradual stretching of the V-O-V bond proves that more and more NBOs are formed. The stretching in the planar direction and substitution of Ba$^{2+}$ with Li$^+$ affects axial V=O bond whose bond length reduces a little. So, the distorted VO$_6$ octahedron gets stretched in the direction perpendicular to the axial bond with the addition of more Li$_2$O which reduces the tetragonal distortion of the VO$_6$ octahedra. This result is supported by the EPR analysis, where the ratio $\frac{\Delta g_{\parallel}}{\Delta g_{\perp}}$ continuously decreases from VBL10 to VBL25. The decrease in this ratio shows that the tetragonal distortion of the VO$_6$ octahedra decreases with increasing Li$_2$O content in the planar direction. So, as we replace BaO in the glasses with Li$_2$O, the environment around VO$_6$ octahedra which is the basic building block in our glasses becomes more and more homogenous.

However, the concentration of $V^{4+}$ shows a non-monotonic variation with an increase in $Li_2O$ content similar to the non-monotonic variation in the peak area of $V^{4+}=O$ bond in the Raman study which indicates an increase in the number of lithium atoms sitting near the double bond.

Thus, we show that EPR and Raman analysis corroborate perfectly to envisage the structural changes in the studied lithium substituted barium vanadate glasses.

## 5. Acknowledgments:

PG is grateful to the Principal and management of St. Joseph's College for their support in her doctoral work. PG is thankful to DST for the DST FIST fund (SR/FST/College-324/2016) awarded to St. Joseph's College and acknowledges the use of the furnace and oven bought under it. GVH is thankful to VGST for awarding a research grant under KFIST L2 GRD No. 648(2017-18). We gratefully acknowledge the support given by Prof. K.P. Ramesh (IISc) to use the facilities at the physics department of the Indian Institute of Sciences, Bangalore. We acknowledge the use of the Sophisticated Analytical Instrument Facility (SAIF), IIT Bombay for obtaining the EPR spectra of our glass samples and thank the facility in charge for her help.